\documentclass[12pt]{iopart}
\usepackage{times}

\newcommand{\xZ}{Z\!\!\!Z}
\newcommand{\xR}{I\!\!R}

\newcommand{\xbe}{\begin{equation}}
\newcommand{\xbea}{\begin{eqnarray}}
\newcommand{\xeea}{\end{eqnarray}}

\begin{document}

\title{$\xZ_n$-Graded Topological Generalizations of Supersymmetry and Orthofermion Algebra}
\author{A Mostafazadeh}
\address{Department of Mathematics, Ko\c{c} University, Rumelifeneri Yolu, 34450 Sariyer, Istanbul, Turkey}
\begin{abstract}
We review various generalizations of supersymmetry and discuss their relationship. In particular, we show how supersymmetry, parasupersymmetry, fractional supersymmetry, orthosupersymmetry, and 
the $\xZ_n$-graded topological symmetries are related.
\end{abstract}

\section{Introduction}

The advent of supersymmetric quantum mechanics (SQM) in the 1980s \cite{witten-82} and its remarkable applications \cite{susy} have since motivated many researchers to seek for generalizations of SQM. Most of these generalizations are algebraic in nature in the sense that they are defined in terms of an operator algebra involving a central element called the Hamiltonian $H$ and a number of noncentral operators ${\cal Q}_a$ and ${\cal Q}_a^\dagger$ called the symmetry generators such that this operator algebra generalizes the algebra of SQM, namely \cite{nicolai,witten-82}
	\xbe
	{\cal Q}_a^2={\cal Q}_a^{\dagger 2}=0,~~~~~[{\cal Q}_a,H]_-=0,~~~~~~
	[{\cal Q}_a,{\cal Q}_b^\dagger]_+=2\delta_{ab}H,
	\label{sqm}
	\end{equation}
where $[A,B]_\pm:=AB\pm BA$ and $a,b=1,2,\cdots,{\cal N}$ for some ${\cal N}\in\xZ^+$. This
is known as the algebra of type $N=2{\cal N}$ supersymmetry (SUSY). In what follows we shall only consider the case ${\cal N}=1$ and drop the label $a=1$. The general case (${\cal N}>1$) may be treated similarly. 

An algebraic generalization of the $N=2$ SUSY corresponds to an associate operator algebra involving $H$, ${\cal Q}$, ${\cal Q}^\dagger$, and a set of more general defining relations. Typical examples are Parasupersymmetry (PSUSY), fractional supersymmetry (FSUSY), and orthosupersymmetry (OSUSY). Note that in ordinary unitary quantum mechanics, one demands $H$ to be Hermitian. Therefore, one considers unitary ($*$-representations) of the underlying operator algebra in a Hilbert space.

An alternative approach to generalize SQM is to adopt the grading and the topological properties of SQM as the guiding principle \cite{mpla,npb}. This leads to a set of generalizations of SQM, called the $\xZ_n$-graded topological symmetries (TS), for which one can define certain integer-valued topological invariants that generalize the Witten index of SUSY
\cite{witten-82}. 

The aim of this article is to provide a brief review of the origins of the above mentioned generalizations of SQM and their relationships.

\section{SUSY Algebra and Its Statistical Generalizations}

Consider the Hamiltonian of the Bose-Fermi oscillator \cite{bryce}
	\xbe
	H:=N_++N_-,
	\label{H}
	\end{equation}
where $N_\pm:=a_\pm^\dagger a_\pm$ is the number operator for a bosonic or fermionic degree of freedom depending on whether its subscript is $+$ or $-$ respectively, and $a_\pm$
is the corresponding annihilation operator. Then one can check that $H$ together with ${\cal Q}:=\sqrt 2 a_+^\dagger a_-$ satisfies the $N=2$ SUSY algebra~(\ref{sqm}). Here one uses the algebraic identities of Bose and Fermi statistics. If one identifies the subscript $-$ with a parafermion \cite{green} of order $p$ and assumes relative bosonic statistics \cite{greenberger}, then $H$ and 
${\cal Q}:=a_+^\dagger a_-$ satisfy the algebra of PSUSY of order $p$, \cite{psqm}, 
	\xbe
	{\cal Q}^{p+1}=0,~~~~~[{\cal Q},H]_-=0,~~~~~~
	\sum_{k=0}^p {\cal Q}^{p-k}{\cal Q}^\dagger {\cal Q}^k=2p{\cal Q}^{p-1}H.
	\label{psqm}
	\end{equation}
Similarly if $a_\alpha$, with $\alpha=1,2,\cdots,p$, denote the annihilation operators associated with an orthofermion of order $p$, $H$ and ${\cal Q}_\alpha:=a_+^\dagger a_\alpha$ satisfy the algebra of OSUSY of order $p$, \cite{mishra}, 
 	\xbe
	{\cal Q}_\alpha {\cal Q}_\beta=0,~~~~~[{\cal Q}_\alpha,H]_-=0,~~~~~
	{\cal Q}_\alpha {\cal Q}_\beta^\dagger+\delta_{\alpha\beta}\sum_{\gamma=1}^p
	{\cal Q}_\gamma^\dagger {\cal Q}_\gamma=2\delta_{\alpha\beta}H.
	\label{osqm}
	\end{equation}
Note that both (\ref{psqm}) and (\ref{osqm}) reduce to the $N=2$ SUSY algebra for $p=1$. Therefore, PSUSY and OSUSY are generalizations of SUSY. Next consider the operator algebra for the FSUSY of order $F=2,3,\cdots$, namely ${\cal Q}^F=H$, \cite{fsqm}. Again if we set $F=2$ we obtain the algebra of $N=1$ SUSY for which ${\cal Q}={\cal Q}^\dagger$. Therefore FSUSY is also an algebraic generalization of SUSY.

\section{Topological Properties of SUSY and PSUSY of Order 2}

Given a supersymmetric quantum system, one can show that the difference of the number of zero-energy bosonic and fermionic states, which is called the Witten index, remains invariant under arbitrary SUSY-preserving continuous deformations of the system \cite{witten-82}. This observation serves as the basis of the supersymmetric proofs of the celebrated Atiyah-Singer index theorem, 
\cite{index-thm}. The fact that SUSY has a rich topological content raises the natural question whether its generalizations share similar properties. 

In order to define the Witten index for a general supersymmetric system, one uses the double ($\xZ_2$) grading of its Hilbert space. The Hilbert space ${\cal H}$ always admits a grading operator $\tau$ fulfilling $\tau=\tau^\dagger=\tau^{-1}$ and $[\tau,{\cal Q}]_+=0$. This operator splits 
${\cal H}$ into the direct sum of its two eigenspaces ${\cal H}_\pm$. The elements of ${\cal H}_+$ and ${\cal H}_-$ are respectively called `bosonic' and `fermionic' state vectors. We shall instead use the term: `vectors with definite grade  $+$ or $-$'. Another important ingredient that one uses to establish  topological invariance of the Witten index is the particular spectral degeneracy structure (SDS) of supersymmetric systems: the energy spectrum is nonnegative, and positive-energy eigenstates come in degenerate pairs with opposite grade. The proof of the fact that every supersymmetric system has this particular SDS is equivalent to the problem of finding unitary irreducible representation (irrep) of the SUSY algebra~(\ref{sqm}). It turns out that there are only two types of irreps namely the trivial 1-dim.\ irrep in which $H={\cal Q}=0$ and the  2-dim.\ irreps labeled by $E\in\xR^+$ in which $H=EI$,  $I$ is the $2\times 2$ identity matrix, $\tau=\sigma_3$, ${\cal Q}=\sqrt{E/2}(\sigma_1-i\sigma_2)$, and $\sigma_i$ are Pauli matrices.

In \cite{esqm} the authors explored the topological invariants of the extended and generalized SUSY. These are essentially SUSYs with more than one grading operator. The first study of the topological properties of a genuine generalization of SUSY is \cite{ijmpa-96} where the question of defining an analog of the Witten index for PSUSY of order 2 was addressed. Again this question may be reduced to finding unitary irreps of the algebra of PSUSY of order 2. It turns out that there are three types of irreps: the trivial 1-dim.\ irreps labeled by $E\in\xR$ in which ${\cal Q}=0$ and $H=E$, the 2-dim.\ irreps that coincide with those of the SUSY algebra, and the 3-dim.\ irreps labeled by 
$E\in\xR^+$ and $t\in[-1,1]$ in which $H=EI$, $I$ is the $3\times 3$ identity matrix, and
	\xbe
	\tau=\sqrt{E}\left(\begin{array}{ccc}
	1 & 0 & 0\\
	0 & 1 & 0\\
	0 & 0 & -1\end{array}\right),~~~~~
	{\cal Q}=\sqrt{E}\left(\begin{array}{ccc}
	0 & 0 & i(1+t)\\
	0 & 0 & \sqrt{1-t^2}\\
	i(1-t) & \sqrt{1-t^2} & 0\end{array}\right).
	\label{irrep}
	\end{equation}
One can use the above results on the representation theory of the PSUSY algebra of order 2 to
infer that only for a special type of PSUSY of order 2 one can define a topological invariant. These correspond to the systems with nonnegative spectrum whose positive energy eigenvalues are all triply degenerate. These systems have been studied and classified in \cite{ijmpa-97}. The extension of the results of \cite{ijmpa-96,ijmpa-97} to PSUSY of order $p>2$ has not been possible mainly due to difficulties associated with the representation theory of the algebra~(\ref{psqm}) for $p>2$.

\section{Topological Generalizations of SUSY}

The study of the topological properties of SUSY and PSUSY of order 2 reveals the fact that the basic ingredients responsible for these properties are their grading and degeneracy structures. This leads to the point of view that one should define a set of generalizations of SUSY by requiring that they possess appropriate grading and degeneracy structures. By definition these symmetries involve a set of integer-valued topological invariants and include SUSY as a special case. They are consequently called topological symmetries (TS). TSs were initially defined for systems with a $\xZ_2$-graded Hilbert space in \cite{mpla}. Their $\xZ_n$-graded generalization was subsequently considered in \cite{npb}. 

$\xZ_n$-graded TSs are described by $n$ positive integers $m_1,\cdots,m_n$ as follows:
1.~The Hilbert space ${\cal H}$ is $\xZ_n$-graded, i.e., there are (nonzero) subspaces ${\cal H}_\ell$ such that ${\cal H}={\cal H}_1\oplus\cdots\oplus{\cal H}_n$. The elements of ${\cal H}_\ell$ is said to have (definite) grade $\ell$; 2.~The Hamiltonian maps ${\cal H}_\ell$ to ${\cal H}_\ell$; 3.~The energy spectrum is nonnegative; 4.~For every positive energy eigenvalue $E$, there is a $\lambda_E\in\xZ^+$ such that the eigenspace of $E$ is spanned by $\lambda_Em_1$ vectors of grade $1$, $\lambda_Em_2$ vectors of grade $2$, $\cdots$, and $\lambda_Em_n$ vectors of grade $n$. Given this definition one can easily show that the integers $\Delta_{ij}:=m_in^{(0)}_j-m_jn^{(0)}_i$, with 
$n^{(0)}_\ell$ denoting the number of zero-energy states of grade $\ell$, are topological invariants. 

It is quite remarkable that the rather general definition of TS is indeed sufficient to determine the underlying operator algebra. It turns out that the algebras of SUSY, PSUSY of order 2, and FSUSY of 
arbitrary order are among the algebras supporting TSs. In particular, for $n=2$ and $m_1=m_2=1$ one derives a unique operator algebra that is identical with the SUSY algebra~(\ref{sqm}) with $N=2$. Similarly, for $n=m_1=2$ and $m_2=1$ one obtains the PSUSY algebra~(\ref{psqm}) with $p=2$. Finally, for arbitrary $n$ and $m_1=m_2=\cdots=m_\ell=1$ one obtains an algebra defined by the
FSUSY relation $H={\cal Q}^{n+1}$ together with a couple of additional relations, \cite{npb}.

\section{Orthofermion Algebra and Topological Symmetries}

In \cite{jpa-01}, it has been shown that the orthofermion algebra of order $p$, \cite{ofermi}
	\xbe
	a_\alpha a_\beta=0,~~~~~~
	a_\alpha a_\beta^\dagger +\delta_{\alpha \beta} \sum^p_{\gamma=1}
	a_\gamma^\dagger a_\gamma=\delta_{\alpha \beta},
   	\label{ofermi}
	\end{equation}
has a unique nontrivial unitary irrep which has dimension $p+1$, that in this representation $a_\alpha$ is represented by a matrix with entries $[a_\alpha]_{ij}=\delta_{i,1}\delta_{j,\alpha+1}$, 
$i,j=1,\cdots,p+1$, and that every representation of (\ref{ofermi}) is completely reducible to copies of the above irrep and the trivial irrep. Furthermore, one can check that the operators $L:=a_1+\sum_{\alpha=2}^p a_{\alpha-1}^\dagger a_\alpha$ and $J:=L+a_p^\dagger$ respectively satisfy $L^{p+1}=0$, $\sum_{k=0}^p L^{p-k}L^\dagger L^k=p L^{p-1}$, and $J^{p+1}=1$. This in turn implies that every system with an OSUSY of order $p$ has a $\xZ_n$-graded TS, a PSUSY of order $p$, and a FSUSY of order $p+1$, \cite{jpa-01,mpla-02}.

For the case $p=3$, one can actually find a realization of this type of TS using Fredholm (possibly differential) operators acting in an inner product (Hilbert) space. This realization has been studied and the relation between the topological invariants $\Delta_{ij}$ and the analytic indices of the associated operators has been discussed in \cite{npb-02}.

\end{document}